\documentclass[twocolumn,twoside,slac_two]{revtex4}
\usepackage{graphicx}
\usepackage{fancyhdr}
\newcommand{\Fermic}{\emph{Fermi}}
\newcommand{\Fermi}{\Fermic\ }
\newcommand{\FermiLATc}{\Fermic~LAT}
\newcommand{\FermiLAT}{\FermiLATc\ }

\begin{document}

\title{Calorimeter-only analysis of the Fermi Large Area Telescope}

%

\author{M. Takahashi}
\affiliation{The University of Tokyo}

\author{R. Caputo}
\affiliation{University of California Santa Cruz}

\author{D. Paneque}
\affiliation{Max Planck Institute for Physics, The University of Tokyo}

\author{C. Sgr\`o}
\affiliation{INFN Sezione di Pisa}

\author{on behalf of the Fermi-LAT collaboration}
\affiliation{}



\begin{abstract}

Above tens of GeV, $\gamma$-ray observations with the \Fermi\ Large Area
Telescope (LAT) can be dominated by statistical uncertainties due to
the low flux of sources and the limited acceptance. We are developing
a new event class which can improve the acceptance: the
``Calorimeter-only (CalOnly)'' event class. The LAT has three
detectors: the tracker, the calorimeter, and the anti-coincidence detector. While the conventional event classes require information from the tracker, the CalOnly event class is meant to be used when there is no usable tracker information. Although CalOnly events have poor angular resolution and a worse signal/background separation compared to those LAT events with usable tracker information,  they can increase the instrument acceptance above few tens of GeV, where the performance of Fermi-LAT is limited by low photon statistics. In these proceedings we explain the concept and report some preliminary characteristics of this novel analysis.

\end{abstract}

\maketitle

\thispagestyle{fancy}


\section{Introduction}

The \Fermi\ Large Area Telescope (\FermiLAT) is an instrument on the \Fermi\ $\gamma$-ray telescope operating from $20$\,MeV to
over $300$\,GeV. The instrument is a $4 \times 4$ array of identical
towers, each one consisting of a tracker--converter 
(TKR), based on Silicon detector layers interleaved with Tungsten foils, 
where the photons have a high probability of converting to pairs, 
which are tracked to allow reconstruction of 
the $\gamma$-ray direction and a segmented calorimeter (CAL), 
made of CsI crystal bars, where the electromagnetic shower is partially absorbed
to measure the $\gamma$-ray energy. The tracker is covered
with an anti-coincidence detector (ACD) to reject the charged-particle
background.  Further details on the LAT, its performance, and calibration 
are given by \cite{FermiMission} and \cite{LATPerformance}.

\begin{figure}
\includegraphics[width=65mm]{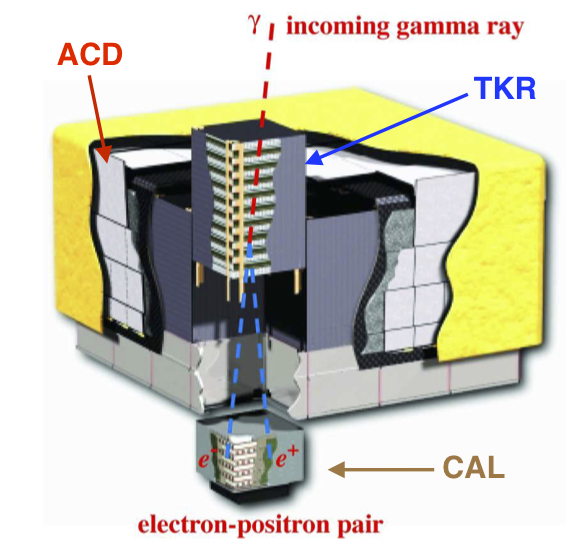}%
\caption{Schematic diagram of the Large Area Telescope. The telescope’s dimensions are $\rm{1.8 m \times 1.8 m \times 0.72 m}$. \cite{FermiMission}}
\label{LATdetecter}
\end{figure}

Most of the science done with \FermiLAT spans photons with energies
from 50 MeV to about 10 GeV, where the sensitivity of the instrument
is good and the available number of detected photons high. 
However, there are many sources which emit $\gamma$-rays above a few tens of GeV. 
These energies that are almost accessible by the current generation
of Imaging Atmospheric Cherenkov Telescopes (IACTs). Even though the detection area of LAT is
small (in comparison to that of IACTs), \FermiLAT provides all-sky
coverage and a very high duty cycle, which are crucial characteristics
for producing $\gamma$-ray source catalogs and study source variability
in an unbiased way. A prime example is the the first \FermiLAT
catalog of $>$10 GeV sources (1FHL) \cite{1FHL}, which contains 514 sources, out
of which $\sim$100 sources have already been detected at very high
energies($>$100 GeV, or VHE),
and $\sim$200 additional sources have been identified as good
candidates to be VHE emitters and be detected with IACTs.

The performance of \FermiLAT above 10 GeV is
excellent. The angular resolution and signal/background separation is
best at the highest photon energies, where one only suffers from a slight
deterioration of the energy resolution due to the fact that the
showers are no longer contained in the calorimeter. However, the
steep falling photon flux with energy of most $\gamma$-ray
sources, together with the relatively small effective area of LAT
($\sim$1m$^2$), results in a substantial limitation due to
the very low number of detected photons (e.g., in the 1FHL, many
sources were characterized with only 4--5 photon events over
a background signal of 0--1 event). 
The low statistics from $\gamma$-ray sources is going
to be an even larger problem for the second \Fermi\ high-energy LAT catalog 
(2FHL, in preparation), which is expected to consist on $\gamma$-ray sources 
detected above 50 GeV (instead of 10 GeV).

In these proceedings we report an analysis which can help
increase the photon statistics at few tens of GeV, hence improving
the ability to perform science at the highest LAT energies, where
the IACTs start operating. The methodology is still being
developed. Here we only present the concept and report some preliminary characteristics.

\section{The Calorimeter-only (CalOnly) Fermi-LAT analysis}

\begin{figure}
\includegraphics[width=85mm]{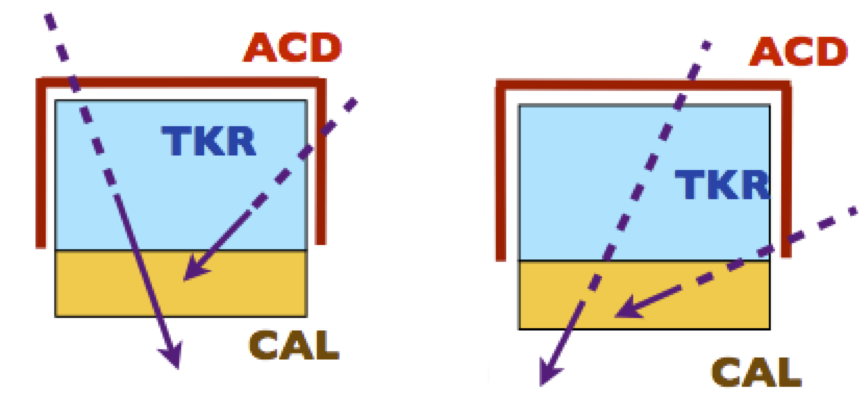}%
\caption{The left panel depicts two events that convert to electron pairs
  in the TKR (from the broken lines to the solid lines in these
  figures). These events would be used in the regular \FermiLAT
  analysis. The right panel shows two events without usable TKR
  information. One is a side-entering event that goes through a small
  fraction of the TKR. The other one crosses the entire TKR, but converts
  to an electron pair in the CAL. These two events would be rejected and
  not be used for regular LAT analysis. These are the type of events
  which could be recovered with a dedicated analysis that does not
  require TKR information, the so-called CalOnly events. }
\label{Event_Tkt_Cal}
\end{figure}

The regular \FermiLAT event classes require usable information from the TKR. 
This is a sensible approach, given
that the TKR information is crucial to determine accurately the incoming
direction of the $\gamma$-ray event. The LAT TKR comprises only $\sim$1.5 radiation lengths (on axis),
 which means that a large fraction of $\gamma$-rays from the astrophysical 
sources are discarded at the very beginning of the analysis because they do
not convert in the TKR,
or they convert in the bottom layers and the TKR information
is not sufficient for a proper determination of the incoming direction
of the $\gamma$-ray event. This situation is depicted in Figure \ref{Event_Tkt_Cal}.

In the standard LAT analysis, the CAL is essentially used for signal/background
separation (together with the TKR and ACD) and to determine the energy
of the $\gamma$-rays and electrons.
The LAT has a hodoscopic calorimeter, consisting of 16 towers with 8 layers of 12 crystals each because of three driving reasons. First, shower profiling improves energy reconstruction. Next, shower topology contains valuable information for signal/background separation. Last, it realizes  independent event acceptance and reconstruction. 
  
To create a usable event class without using TKR information, one must determine the incoming direction of the $\gamma$-rays with sufficient resolution (a few degrees) while keeping a reasonable background rejection ($\sim$0.999).This would increase the number of
available high-energy events for performing $\gamma$-ray astronomy.
The above mentioned (non-standard) LAT analysis, dubbed
Calorimeter-only (CalOnly) analysis, is currently being developed by
the \FermiLAT collaboration, and is aimed to deliver yet another class
of events, the CalOnly event class, which may be added to the other
photon event classes coming from the regular LAT analysis. For the CalOnly analysis, it fully reconstructs the electromagnetic showers and determine its main axis, which points to the direction of the incoming $\gamma$-ray event.

Since the main event trigger for LAT is based on the TKR, those events
with no usable TKR information have a low chance of being recorded
and transferred to the ground. However, large energy depositions in CAL generate a trigger that is fully independent of TKR, and on-board event selection records events that deposit energy larger than 20 GeV in CAL.
Consequently, the CalOnly event class will only be effective
above a few tens of GeV.

\begin{figure*}[t]
\centering
\includegraphics[width=\textwidth]{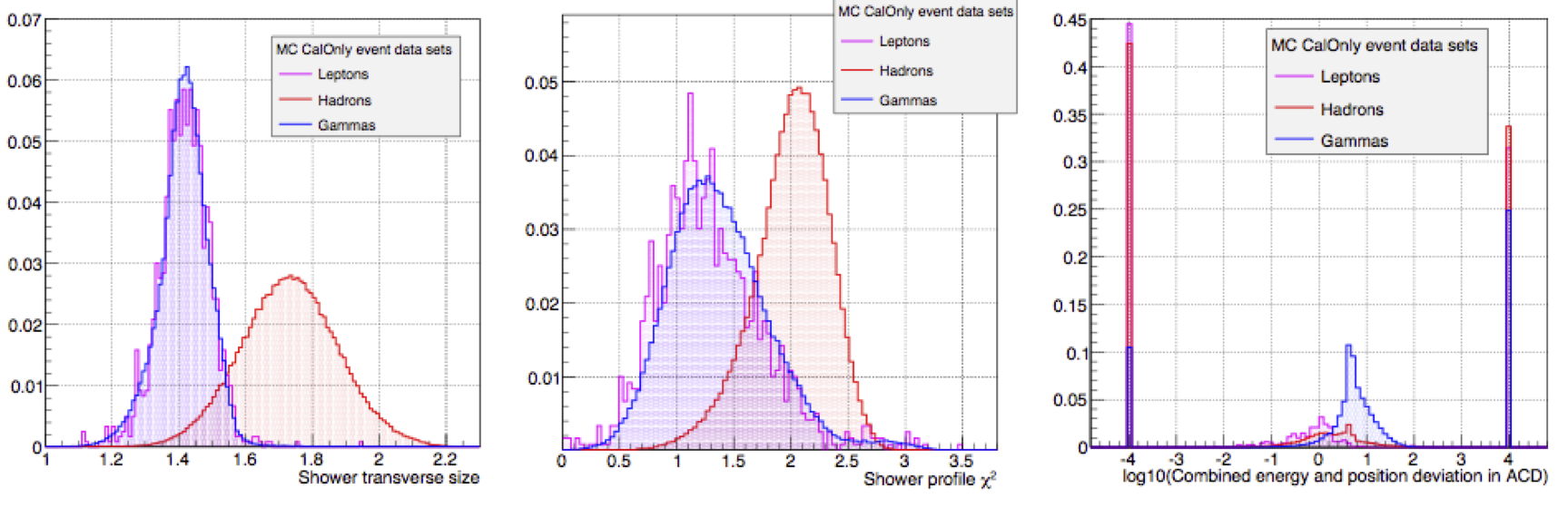}
\caption{The left, middle and right panels show the logarithm of the CAL first cluster transverse profile RMS, the logarithm of $\chi^2$ of the profile fit computed over a 100mm radius cylinder around the trajectory, and the logarithm of combined energy and position deviation in ACD respectably. The last one needs some more description. It is the number of sigmas less than an expected MIP signal, combined with the number of sigmas the track propagation is away from tile or ribbon most likely to veto the first CAL cluster and the number of sigmas the energy deposited in the ACDs is away from the expected amount. If the value of this combined quantity is zero, then the log10 of this quantity is set to -4. This is more likely to happen to MIPs than for $\gamma$-rays. If there are no tracks associated to ACD signals, this quantity is set to +4. Charge particles are likely to have smaller values than the $\gamma$-rays.} 
\label{MVA_variables}
\end{figure*}

\subsection{Pass 8} 

The development of the CalOnly analysis is done in the context of Pass 8,
which is the new iteration of the LAT event-level analysis package.
Pass 8 was originally designed to address the effect of coincidences 
with cosmic rays (`ghost' events), but quickly evolved in a
comprehensive revision of the instrument simulation, the event reconstruction, 
and the background rejection, with the goal of improving all the aspects
of the LAT performance: larger acceptance, better angular and energy resolution,
and extension of the energy reach below 100 MeV and in the TeV range.

The details on the Pass 8 analysis chain can be found in~\cite{Pass8}. 
The CAL reconstruction begins with a clustering stage that tries to isolate 
the genuine $\gamma$-ray shower from smaller energy deposition due to ghost events.
At this point we can exploit the segmentation of the CAL to identify 
the energy deposition centroid and the shower axis (via a moment analysis)
that, for CalOnly, corresponds to the photon incoming direction.
This direction is propagated to the ACD (in addition to the tracks from 
the tracker) in order to associate energy deposition in the tiles
and discriminate charged particle without the TKR direction.
This is one of the main improvements introduced with Pass 8 and proved to be
very useful even if its separation capability is limited by the angular resolution of the CAL, that is obviously worse than that of the TKR..

Another important improvement in the CAL reconstruction is the energy 
measurement that, for CalOnly, is based on a full three-dimensional fit of
the shower energy deposition. This method needs a precise modeling of 
the longitudinal and lateral development of showers inside the CAL and a
reference axis. The latter, usually taken from the tracker, can be obtained 
from the aforementioned moment analysis with a small change in performance.

The last step of the Pass 8 development is the high-level analysis that links 
together all the outputs of the reconstruction and classify events as good 
$\gamma$ rays or not. This is the core of the CalOnly development and 
is described in next section.

\subsection{Signal / Background separation} 

The \FermiLAT needs to reject a cosmic-ray background that outnumbers the
signal ($\gamma$ rays) by many orders of magnitude, and hence an
efficient signal/background separation (rejecting $10^{3-4}$ of the
background events) is required to be able to perform $\gamma$-ray
astronomy.  The LAT background
consists mostly on protons and electrons, but also on alpha
particles and heavy nuclei. 
As the $\gamma$-ray energy increases, we have a natural improvement in
the signal/background ratio due to the fact that most $\gamma$-ray
sources have spectra that can be parameterized with power-law indices
harder than 2.5 (often even harder than 2.0), while the spectra of the
proton background follows a power-law index of $\sim$2.7 and that of the
electrons a power-law index of $\sim$3.1. 

The rejection of the background cosmic-ray events in the CalOnly
analysis is based on the different topology of electromagnetic and
hadronic showers, and the ACD signals produced by the charged
particles. It is worth noting that, while protons and heavy nuclei can
be effectively distinguished from $\gamma$ rays using only information
from the CAL, the electrons/positrons produce electromagnetic showers
that are essentially identical to those of the $\gamma$ rays, and
hence the information from the ACD is crucial to be able to reject
electrons. The left and middle panels in Figures \ref{MVA_variables} show 
the normalized MC distributions of two CAL-related parameters that can effectively distinguish between
electromagnetic ($gammas$ and leptons) and hadronic showers. 
In order to be able to reject the leptons, one needs the help of
ACD-related parameters, as depicted in the right panel in Figure \ref{MVA_variables}.

\begin{figure}[t]
\includegraphics[width=75mm]{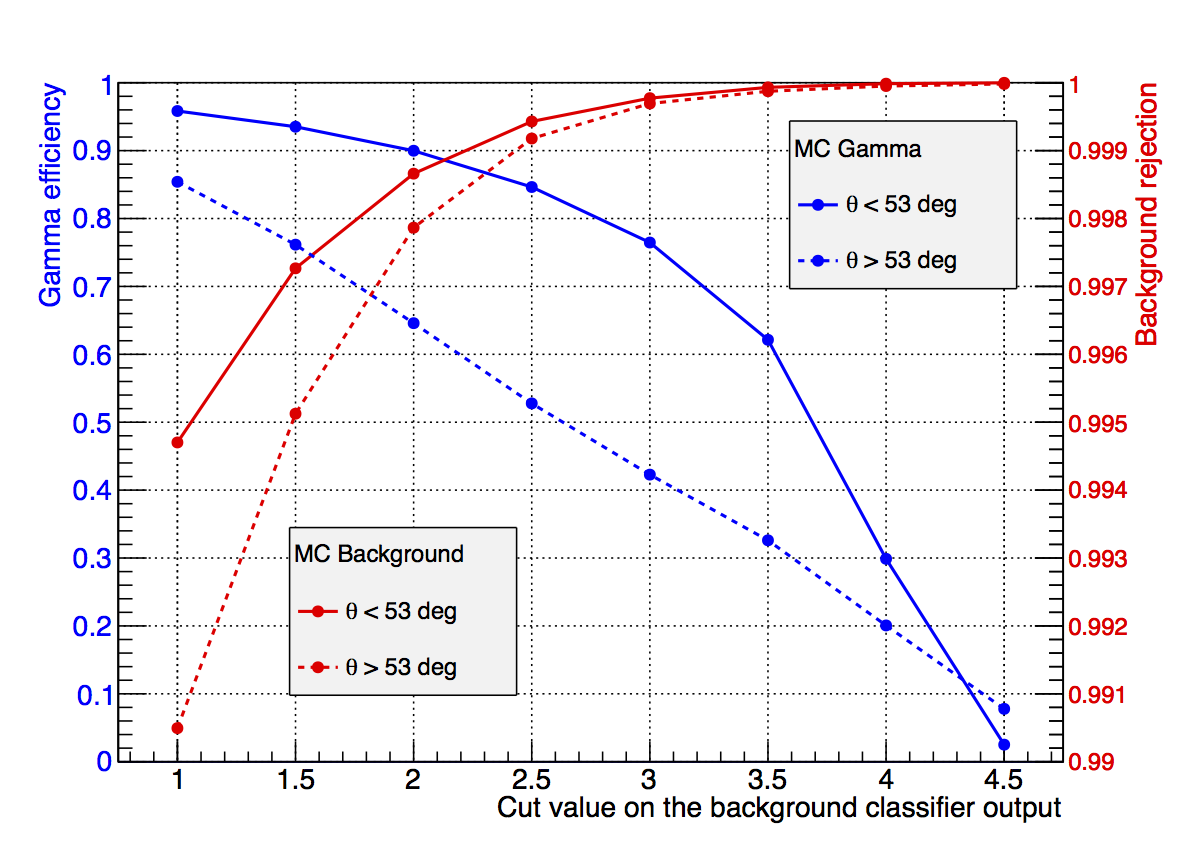}%
\caption{Efficiency of MC $\gamma$ events (blue) and rejection of MC background events (red) vs. cut value of a classifier output.}
\label{Acc_vs_Rej}
\end{figure}

In order to maximize the separation of signal and background, instead
of making simple cuts in the distributions of CAL and ACD parameters
as the ones shown above, we perform the analysis through a multi-variate analysis
(MVA) that uses a large number of CAL and ACD parameters. 
For the most effective background rejection, we applied the Boosted
Decision Tree (BDT), one of the methods of multi-variate analysis. In
this method, we train many classification trees with Monte Carlo (MC) data,
for which can identify unambiguously what is signal and what is
background. 
We are using the ROOT-based TMVA package to train the classification
tree analysis~\cite{TMVA}. 
We build the trees and then evaluate the gamma-likeness of each
event. Next we  can cut on the gamma-likeness and get the events
classified as signal or background. By selecting events with a very
high gamma-likeness, one can increase the purity of the selected data
set, but at the expense of reducing the number of $\gamma$-ray
candidates.  This is represented in Figure.\ref{Acc_vs_Rej}. The
optimal value to increase the signal/background is typically obtained
for a cut value between 2 and 3 (dependent on the energy range and
incidence angle of the $\gamma$-ray considered).

At the present time we are optimizing the classification tree by
modifying the input CAL and ACD parameters (including creating new
composite variables), as well as by building the
classification trees in different modes. Consequently, the results
presented in this proceedings should be considered as preliminary, and
likely reporting a lower limit of the actual performance of the
CalOnly analysis.

\subsection{Quality of the reconstructed events} 
In this section we address the quality of the reconstructed CalOnly
events (after signal/background separation) using dedicated MC
simulations of  $\gamma$-ray events.

Two basic quantities are being evaluated: the angular and the energy
resolution. Given that the thickness of the calorimeter increases
rapidly with the incidence angle, one expects a
different performance for low and high incidence angle $\gamma$ rays.
In this section we define low (high) incidence angle as smaller (larger) than
53 degrees ($\cos(53^{\circ}) \sim 0.6$) and evaluate the performance 
for these two cases. 
And naturally, as it occurs in the regular
\FermiLAT analysis, the performance can also vary with the energy of
the incoming $\gamma$ ray. Here we define low (high) energy as being
in the range $\sim$30-100 GeV ($\sim$100-300 GeV), and evaluate the
performance for these two energy bands. 

\begin{figure}[htb]
\includegraphics[width=75mm]{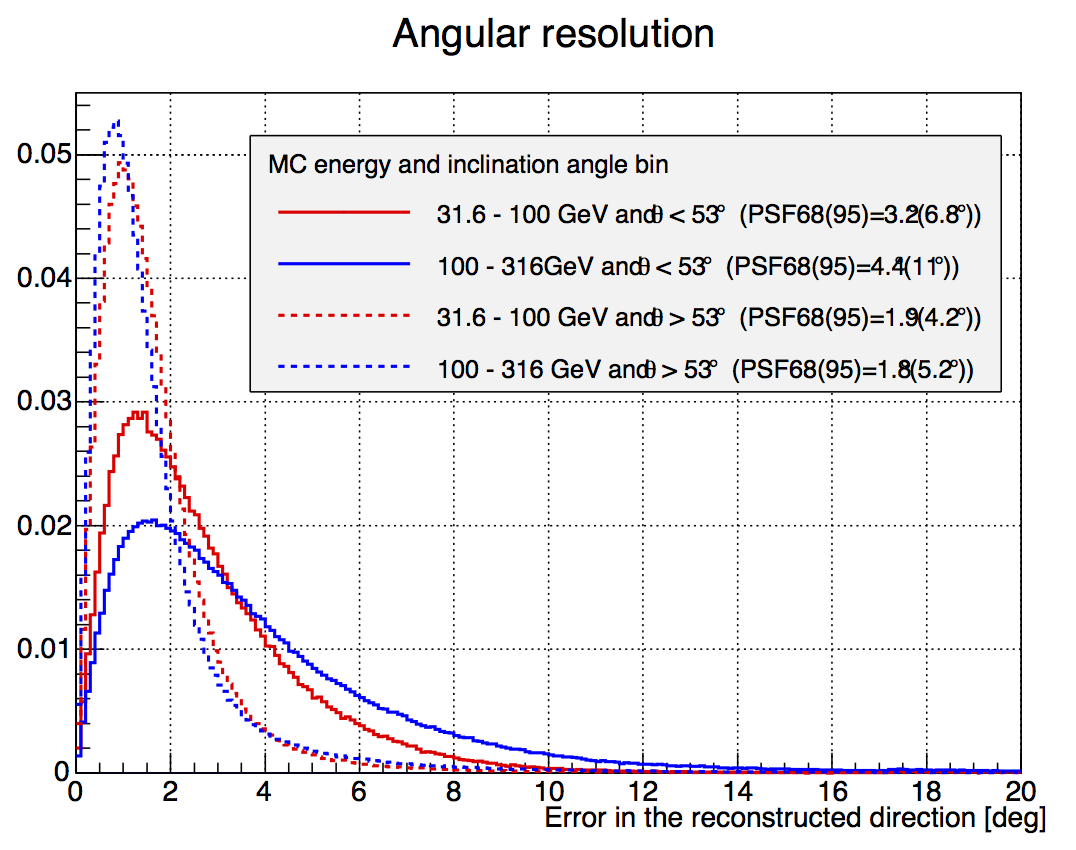}%
\caption{Normalized histograms of the angular distances between reconstructed and MC direction of CalOnly events in 2 inclination angle $\times$ 2 energy bins.}
\label{AngularRes}
\end{figure}

Figure \ref{AngularRes} shows the normalized distributions in the error of the
reconstructed directions for low/high energy bands and incidence
angles. The angular resolution can be defined as the 68\% containment
in those distributions (PSF68), which would result in $\sim$2 degrees
for high inclination $\gamma$ rays, and $\sim$3--4 degrees for low
inclination $\gamma$ rays (with a relatively small dependence on the energy).
The PSF68 for regular LAT photons (i.e. with usable TKR information)
at these energies is $\sim$0.1--0.2 degrees, which is more than one
order of magnitude better than for CalOnly photons.

\begin{figure}[ht]
\includegraphics[width=75mm]{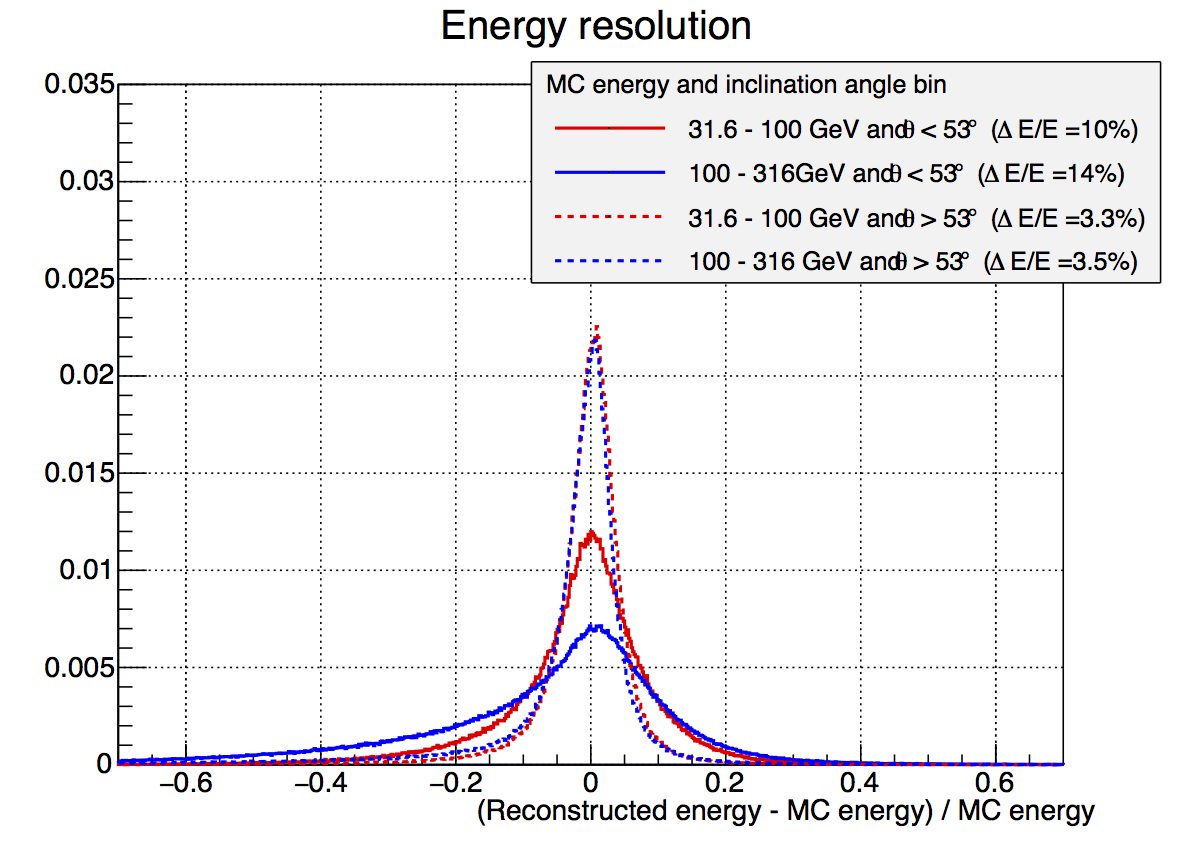}%
\caption{Normalized histograms of the dispersions in the reconstructed energy of MC CalOnly events in 2 inclination angles $\times$ 2 energy bins.}
\label{EnergyRes}
\end{figure}

Figure \ref{EnergyRes} shows the normalized distributions in the error
of the reconstructed energies for low/high energy bands and incidence
angles. The energy resolution can also be defined as the 68\%
containment on these distributions, and using the largest distance
from the peak position to the edge of the 68\% containment.
The energy resolution is $\sim 3-4\%$ for high inclination and $\sim
10-15\%$ for low inclination $\gamma$-rays, with little dependence on
the energy of the event. This performance is very close to that of regular LAT photons.
The quality of the energy reconstruction 
is mainly connected to the path length of shower axis (related to the shower 
containment) and the accuracy of the shower direction reconstruction. 
While the latter is worse for CalOnly events, this class can benefit from a 
larger field of view and therefore longer trajectories.
It must be noted that both direction and energy resolution can be improved 
with a dedicated selection of good quality events, at the price of a lower 
effective area. The best trade off between these conflicting requirements is still to be evaluated.

\section{Conclusions}

Pass 8 provides an unprecedented framework to develop an analysis that
uses events without usable TKR information. 

The CalOnly event class, currently under development within the
Fermi-LAT collaboration, could be used to increase the acceptance of
\FermiLAT above few tens of GeV (where the performance is photon
statistics limited), by recovering for astronomical studies $\gamma$-ray events
without usable TKR information. This implies that CalOnly events will 
have a worse signal/background separation and angular resolution, 
when compared to the regular LAT events. 
However, they can have a better energy resolution,
if considering the high incidence angle events.

The CalOnly event class may be particularly relevant in the following two scientific topics:

\begin{itemize}
\item Search for line-signals potentially coming from Dark Matter annihilation (because of the larger number of events and the excellent energy resolution for the large-incident angle events)
\item Study of transient events like GRBs and AGN flares (because of the larger number of events and the valuable increase in the temporal coverage of the source)
\end{itemize}

\bigskip 
\begin{acknowledgments}
The \textit{Fermi}-LAT Collaboration acknowledges support for LAT development, operation and data analysis from NASA and DOE (United States), CEA/Irfu and IN2P3/CNRS (France), ASI and INFN (Italy), MEXT, KEK, and JAXA (Japan), and the K.A.~Wallenberg Foundation, the Swedish Research Council and the National Space Board (Sweden). Science analysis support in the operations phase from INAF (Italy) and CNES (France) is also gratefully acknowledged.

We would like to thank M. Wood for developing a python module for
multi-variate analysis, which was heavily used in this study. 

\end{acknowledgments}

\bigskip 

\end{document}